\documentclass[11pt]{article}
\usepackage{geometry}                
\usepackage{graphicx}
\usepackage{amssymb}
\usepackage{epstopdf}
\usepackage{authblk}
\usepackage{bibentry}
\usepackage{amsmath, amsthm, amssymb}
\usepackage[numbers]{natbib}
\title{Non-monotonic hydrodynamic lift force on highly-extended polymers near surfaces}
\author[1]{Charles E. Sing}
\author[1]{Alfredo Alexander-Katz \footnote{aalexand@mit.edu}}
\affil[1]{Department of Materials Science and Engineering, Massachusetts Institute of Technology, 77 Mass. Ave. Cambridge MA 02139}

\begin{document}

\begin{abstract}
The hydrodynamic lift force that polymers experience near boundaries is known to be a crucial element when considering rheological flows of dilute polymer solutions. Here we develop theory to describe the hydrodynamic lift force on extended polymers flowing near flat surfaces. The lift force is shown to display a non-monotonic character increasing linearly with the distance to the wall $Z$ in the near-surface regime defined as $Z<L$, with $L$ being the contour length of the polymer. At heights $Z \sim L$ the lift force displays a maximum, and for $Z>L$ we recover the well known far-field result in which the force decays as $Z^{-2}$. Our analytical theory has important implications in understanding adsorption, desorption, and depletion layers of highly extended objects in flow.
\end{abstract}

\maketitle

There is a "depletion zone" in flowing solutions of macromolecules which is characterized by a drastic reduction in the concentration of the polymer near a surface.\protect\cite{Fang:2007p7340,Ma:2005p5986, Jendrejack:2004p2674,CriadoSancho:2000p2666} Over the past decade it has become clear, after early work by Jendrejack, et al., that this behavior is due to a hydrodynamic lift force, which is a direct consequence of polymer stretching under the influence of an external fluid flow.\protect\cite{Ma:2005p5986,Hoda:2007p16,Jendrejack:2004p2674,Jendrejack:2002p824,Hoda:2007p7378,HernandezOrtiz:2006p7379, JendrejackdePablo} The origin of this lift-force arises from a no-slip boundary condition at a solid surface that breaks the fluid flow symmetry around a chain under tension, and the chain feels a resulting upwards force. This force has been widely investigated and applied to a variety of contexts, in particular the description of the aforementioned depletion zone and the case of polymer adsorption.\protect\cite{Ma:2005p5986,Hoda:2007p16,Jendrejack:2004p2674,Jendrejack:2002p824,HernandezOrtiz:2006p7379,Sendner:2007p1461,Hoda:2007p15,Hoda:2007p7378,Sendner:2008p6216,Serr:2009p6217, NETZ, Watari} Ma and Graham were the first to consider this behavior in the context of adsorption, and showed how a dumbell polymer model could be developed that displayed the qualitative characteristics seen in experiment.\protect\cite{Ma:2005p5986} This model was further developed by Hoda and Kumar who extended this model to incorporate the effects of electrostatic interactions.\protect\cite{Hoda:2007p7378} Theoretical and simulation work by Sendner and Netz has further investigated this phenomenon in the context of more refined bead-spring models that more explicitly incorporate the correct hydrodynamics.\protect\cite{Serr:2009p6217,Sendner:2008p6216,Sendner:2007p1461, NETZ} Despite these advances in the theory, comparison of many of these results with experiment reveal qualitative, but not quantitative agreement. Larson, et al. has indicated that the depletion zone near a surface is considerably smaller than predicted by theory, by as much as an order of magnitude.\protect\cite{Fang:2007p7340}

In this Letter we provide a general theory that describes the hydrodynamic lift behavior of a polymer at an arbitrary distance from the surface. Our theory provides an explanation of the discrepancies seen between experiment, simulation, and theory for surface depletion, and also an understanding of the lift forces that are relevant for polymer adsorption and desorption. In particular we note that previous models addressing this force, such as the dumbbell model, uses an explicit far-field assumption, the validity of which has been questioned by Hernandez-Ortiz, et al. and Hoda, et al.. However, a complete quantitative description of this regime has not yet been developed.\protect\cite{HernandezOrtiz:2006p7379,Hoda:2007p7378} Here, we provide analytical results that capture the fundamental physical picture in both the far-field and near-surface regimes. Our analytical theory supposes no underlying chain model, and relies only on geometric parameters and flow conditions. It shows for the first time that the lift force on an extended object, like an extended polymer, can become non-monotonic, increasing linearly with the distance from the surface. In the far-field regime we recover the well known results that the force decreases quadratically with the distance. Of further interest is the fact that the crossover from the near-surface to the far-field regime where the lift force is maximal occurs at a height $Z^* \sim L$, where $L$ corresponds to the polymer extension length. Since $L$ can be extremely large, particularly in strong flows, the near field regime may extend far away from the surface.  

\begin{figure}
\includegraphics[width=150mm]{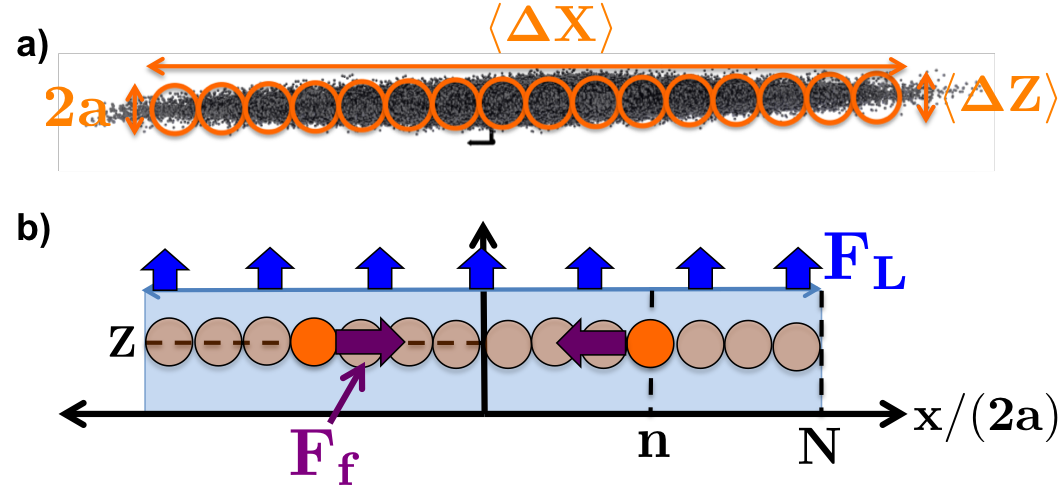}
\caption{(a) Simulation data showing the accumulation of bead positions over 4800 $\tau$ at $\dot{\gamma} \tau= 2.1$ for a $\theta$-chain of $100$ beads. We approximate this geometry using a bead-rod model, which groups portion of the chain into beads (orange circles). This geometry is shown in (b), and has $2N = \langle \Delta X \rangle / \langle \Delta Z \rangle$ beads of radius $a = \langle \Delta Z \rangle/2$ parallel to the surface at height $Z$. Individual dumbbell-pairs feel a net force $F_f$ which is equivalent to the gradient of the tensile force along the chain in the continuum limit. All pairs of dumbbells contribute to an overall lift force $F_L$. \label{Fig1} \vspace{-15pt}}
\end{figure}

To develop this model, we only need to consider the geometry of a polymer chain near a surface. Here we provide a description similar to the blob model of a polymer chain under tension by representing the chain as a series of hydrodynamic beads whose length is equal to the ratio of the average length to the average width of the elongated polymer chain $2N = \langle \Delta X \rangle / \langle \Delta Z \rangle$ (see Figure~\ref{Fig1}a for a schematic).\protect\cite{DeGennes} This description of the polymer chain is amenable to the incorporation of molecular theory; for example a straightforward calculation of this geometry in shear flow based on the work of DeGennes is performed in the Supplemental Material (SM), and yields the relationship $2N = \sqrt{1+ (\dot{\gamma} \tau_Z)^2/(2E^2)}$, where $\dot{\gamma} \tau_Z = Wi$ is the shear rate rendered dimensionless through the use of the chain relaxation time $\tau_Z$ and $E$ is Peterlin's chain stretching parameter that is $1$ at low extension and diverges at high extensions.\protect\cite{DEGENNES:1974p2152} We introduce a single length scale $2a = \langle \Delta Z \rangle$ to describe the chain dimension normal to the wall that effectively corresponds to the bead radius. This consequently allows us to represent our system as a bead-spring chain that lies parallel to the surface, in the same spirit as Sendner and Netz (see Figure~\ref{Fig1}b for a schematic).\protect\cite{Serr:2009p6217,Sendner:2008p6216} We can then determine an analytical form for the hydrodynamic lift force of this fully elongated polymer using the Blake-Oseen tensor, which has been shown to provide a good description of hydrodynamics near surfaces in the regime of interest ($Z > a$).\protect\cite{BLAKE:1974p2245, Karrila} This geometry conveniently enables us to use a simplified version of the hydrodynamic interaction tensor, and captures the essential physics of the lift force in the limit of high stretching. This approach is also the relevant case for a desorbing or adsorbing polymer, which exhibits extended conformations as it interacts with both the external fluid flow and the surface.\protect\cite{Serr:2009p6217,Ladoux:2000p5497} Understanding a polymer in such a limit is also important when considering the behavior of a polymer in channels that are characterized by small length scales.

The lift force on a polymer, or any other deformable component dispersed in a liquid medium near a surface, is due to the hydrodynamic interactions between individual components and the no-slip boundary condition at the surface. The simplest manifestation of the hydrodynamic interaction between an entity and its surroundings is given by the Oseen mobility tensor $\mu_{ij,O}(\mathbf{r}_i, \mathbf{r}_j)$, which describes the entity as a single point force.\protect\cite{DeGennes,BLAKE:1974p2245} This mobility tensor describes the effect of a point force $\mathbf{F}_j (\mathbf{r}_j)$ on the surrounding velocity field $\mathbf{v}_i (\mathbf{r}_i)$ through the equation $\mathbf{v}_i (\mathbf{r}_i) = \mu_{ij,O}(\mathbf{r}_i, \mathbf{r}_j) \cdot \mathbf{F}_j (\mathbf{r}_j)$. To account for the effect of a nearby surface, where a no-slip boundary condition must be maintained, Blake introduced an image system that accounted for the aforementioned boundary condition to produce a new mobility tensor $\mu_{ij,B}$.\protect\cite{BLAKE:1974p2245} See the SM for more discussion of the mobility tensors. 

In our model we consider a finite chain of $2N$ hydrodynamic beads at a distance $Z$ from the surface, each of radius $a$. Using these parameters, we can render all variables dimensionless (rescaled variables denoted by tildes) in terms of distances $a$, energies $kT$, and times $\tau_0 = 6 \pi \eta a^3/(kT)$, where $\eta$ is the solvent viscosity. This geometry is shown schematically in Figure~\ref{Fig1}b. The tension force $F_T$ along the chain is considered to only be in the $x$-direction, such that the chain is in equilibrium in this configuration. Since we are primarily concerned with the lift forces in the direction normal to the surface, we only need to consider the resulting velocity field in the $z$-direction. The relevant component of the mobility tensor is then just $\mu^{xz}_{ij,B}(\mathbf{r_x}) = (-3 \mathbf{r}_x Z^3)/(2 \pi \eta (\mathbf{r}^2_x + 4Z^2)^{5/2})$. To evaluate the overall force,  one can just define the effect of a given dumbell centered at the mid point of the chain that is separated by a distance $4na$ on the velocity field at a dumbell that is separated by a distance $4ma$, where $n$ and $m$ are just indexing parameters (see Fig. 1). The resulting lift force $F_{d}$ from the $n$ dumbbell on the $m$ dumbbell can be writen as:
\begin{eqnarray}
F_{d}(n,m,F_f,\tilde{Z}) 
= -12 \pi \eta a F_f \left[ \mu_{n,m,B}^{zx} + \mu_{n,-m,B}^{zx} \right] \nonumber \\
= \frac{9F_f \tilde{Z}^3}{8} \left[ \frac{(n-m)}{[(n-m)^2 + \tilde{Z}^2]^{5/2}} + \frac{(n+m)}{[(n+m)^2 + \tilde{Z}^2]^{5/2}} \right]
\end{eqnarray}
where $F_f$ is the magnitude of the force on the $n$th bead and the rescaled height $ \tilde{Z}=Z/a$. To determine the overall lift force due to a given dumbell on the entire chain $F_{L,d} (n,F_f,\tilde{Z}$ (including the dumbell itself), we sum over all dumbells $m = 1$ to $N$:
\begin{eqnarray}
F_{L,d} (n,F_f,\tilde{Z})= \sum_{m=1}^N {F_{d}  \approx \int_0^N F_{d} (n,m,F_f,\tilde{Z}) dm} = \nonumber \\
= \frac{3 F_f \tilde{Z}^3}{16} \left[ \left[ (N-n)^2 +  \tilde{Z}^2 \right]^{-3/2} - \left[ (N+n)^2 +  \tilde{Z}^2 \right]^{-3/2} \right]
\label{Sum1}
\end{eqnarray}
where we have used the approximation that $N$ is large enough that we can replace the sum by a continuous integral from $0$ to $N$. To find the total lift force, we perform the summation over all dipoles $n$ and again replace it with an integration:
\begin{equation}
F_L = \sum_{n=1}^N F_{L,d} \approx \int_0^N{ F_{L,d}} dn 
= \frac{3\tilde{Z}^3}{16} \int_0^N{\left( \frac{\partial F_T}{\partial n} \right) \left[ \left[ (N-n)^2 +  \tilde{Z}^2 \right]^{-3/2} - \left[ (N+n)^2 +  \tilde{Z}^2 \right]^{-3/2} \right] dn}
\label{Sum2}
\end{equation}
Since $F_f$, the overall force on the dumbbell, is the difference of the tension on either side of the bead, we make the replacement of this force with its continuum analogue, $(\partial F_T/ \partial n)$. This represents a "kernel" in which the form of the load profile is input into the theory. In this paper we consider a profile that corresponds to shear flows with the extended polymer at a small angle $\theta$ (another example is given in the SM). We incorporate the result $(\partial F_T / \partial n) \approx 6 \pi \eta a^2 \dot{\gamma} \sin{ (2 \theta)} n$. Performing the integration, we get the result:
\begin{equation}
\tilde{F}_{L,shear} = \frac{3}{16}  \dot{\gamma} \tau \sin{ (2 \theta)} \left[ \frac{ 2N^2 \tilde{Z} + \tilde{Z}^3}{\sqrt{4N^2+\tilde{Z}^2}}  - \tilde{Z}^2 \right]
\label{SHEAR}
\end{equation}
For the far-field, we retain the proper $\tilde{Z}^{-2}$ scaling: $\tilde{F}_{L,shear} \approx \frac{3}{8}  \dot{\gamma} \tau \sin{(2 \theta)} \frac{N^4}{\tilde{Z}^2}$, however below a crossover height $\tilde{Z}^* \sim N$ there is a near-surface regime where the lift force becomes non-monotonic and the full equation ~\ref{SHEAR} must be used. This theory can now be related to any microscopic theory of choice, with the geometrical parameters ($N$, $a$, $\tau$) corresponding to chain parameters. We show a straightforward example of this in the SM using Gaussian dumbbells to obtain the well-known far-field scaling relationship for the depletion length, $Z_{dep} \sim Wi^{2/3} n^{1/2}$.\protect\cite{Ma:2005p5986, Jendrejack:2004p2674,Hoda:2007p7378}

\begin{figure}
\includegraphics[width=150mm]{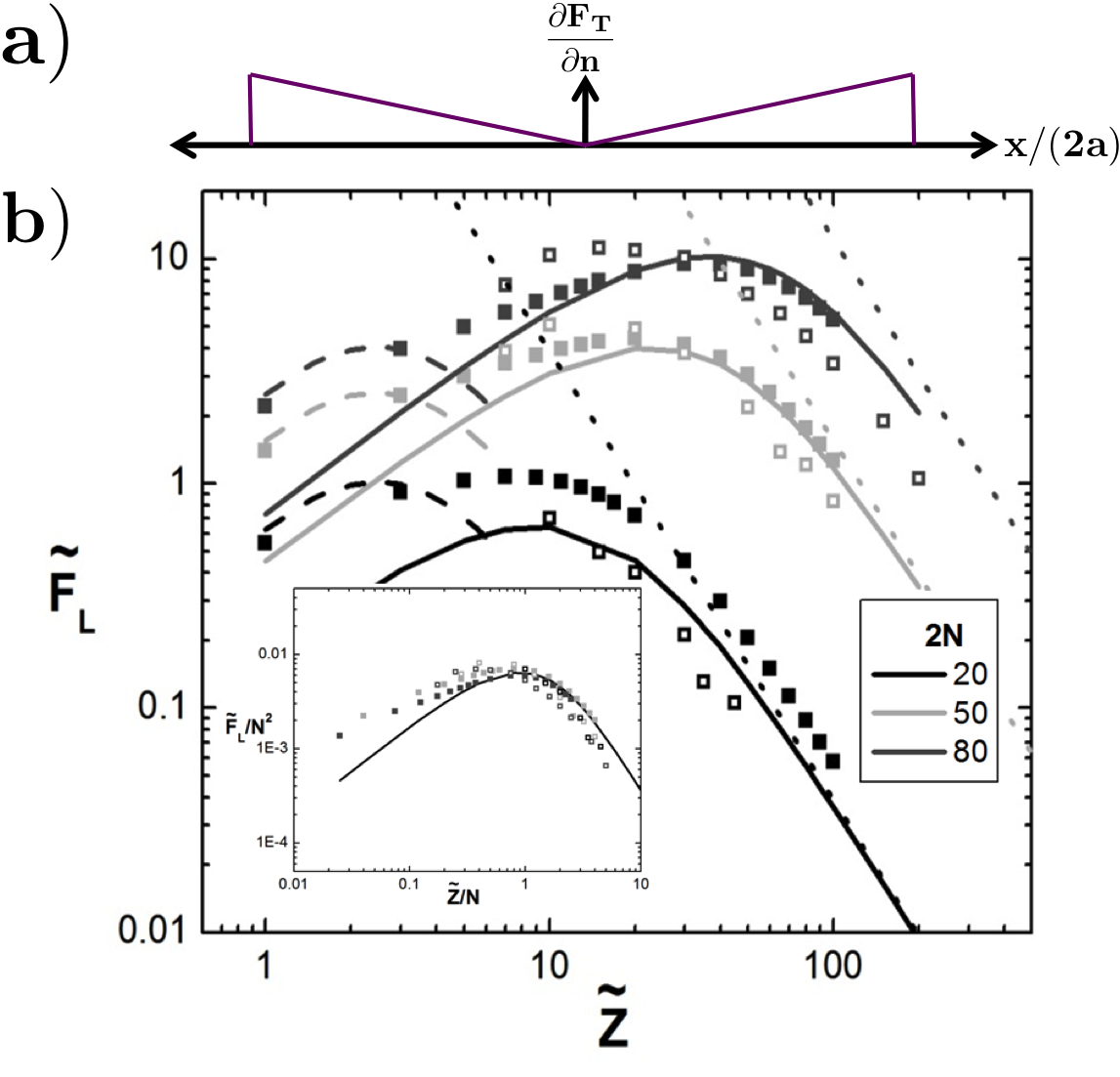}
\caption{(a) Load profile that corresponds to a chain in shear flow, which is placed into the equation~\ref{Sum2} to yield equation~\ref{SHEAR}. It is described by $\tilde{\mu} (\partial \mathbf{\tilde{F}_{f} (\tilde{r}_i)}/\partial n) = \dot{\gamma} \tau \sin{2\theta} (\mathbf{\tilde{r}_{i,x}} - \sum_j^{2N}{\mathbf{\tilde{r}_{j,x}}}/(2N))$. (b)Graph of $\tilde{F}_L$ versus $\tilde{Z}$ for a shear load profile. Solid lines show theoretical results, dashed lines show the effect of bead-discretization at low values of $\tilde{Z}$ (see SM), and dotted lines demonstrate the far-field dumbbell results. Simulation data is also shown, with filled symbols representing data without fluctuations and open symbols representing data including fluctuations. These results collapse onto a single curve (inset) using the scaling $\tilde{Z} \rightarrow \tilde{Z}/N$ and $\tilde{F} \rightarrow \tilde{F}/N^2$. \label{Fig2} \vspace{-15pt}}
\end{figure}

Or results are confirmed by Brownian Dynamics simulations (details in the SM) on chains that are extended parallel to a surface. These bead-spring models represent the effective blobs that make up an extended polymer chain. For these simulations, we incorporate the appropriate hydrodynamics by using the Rotne-Prager-Blake tensor and consider the chain with both the absence and presence of thermal fluctuations.\protect\cite{ROTNE:1969p5862, Karrila} For these simulations we use the force loading profile that corresponds to shear flow, which is given by $\tilde{\mu} (\partial \mathbf{\tilde{F}_{f} (\tilde{r}_i)}/\partial n) = \dot{\gamma} \tau \sin{2\theta} (\mathbf{\tilde{r}_{i,x}} - \sum_j^{2N}{\mathbf{\tilde{r}_{j,x}}}/(2N))$ (shown in Figure~\ref{Fig2}a).

\begin{figure}
\includegraphics[width=75mm]{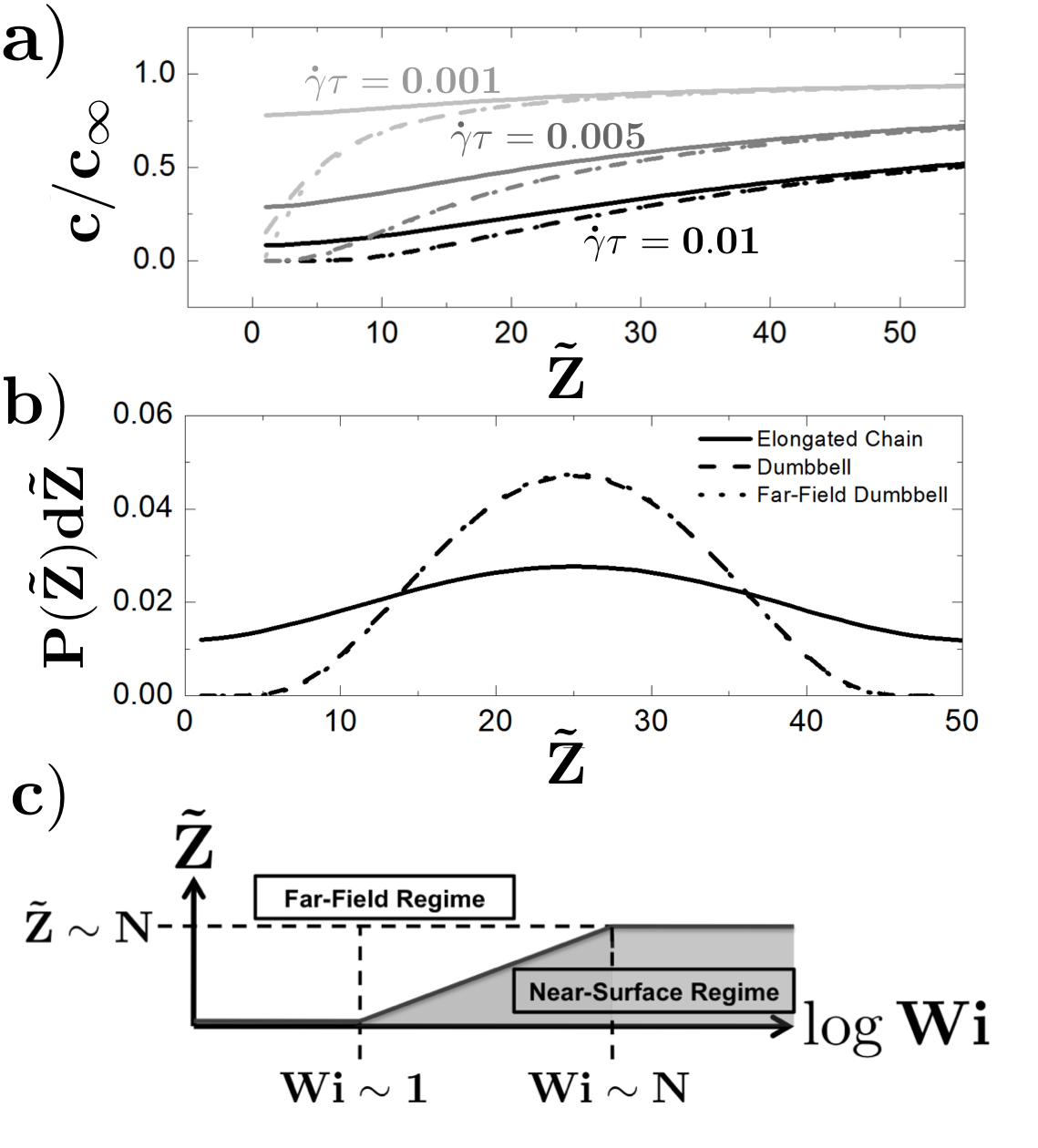}
\caption{(a) Concentration profiles highlighting the difference between the elongated chain model (solid lines), the dumbbell model (dashed lines), and the far-field dumbbell model (dotted lines) for a number of different shear rates near a surface with $N=10$ (given in terms of geometric parameters). The depletion region is often considerably smaller when near-surface hydrodynamics and elongated chain geometries are considered. While direct comparison of these results to Weissenberg numbers is dependent on the specific chain model, a simple scaling model (see SM) using these values suggests they correspond to $Wi \sim 1-10$. All three cases correspond to the same far-field behavior. (b) Probability distribution function for chains in a sheared slit flow. The near-surface hydrodynamics, which are considered in the elongated chain model, have the effect of greatly decreasing surface depletion in channel widths on the order of the chain contour length. $\dot{\gamma} \tau_{eff} = 0.01$ and $N=10$, roughly corresponding to $Wi \sim 5-10$. All three cases correspond to the same far-field behavior. Finite concentrations at $\tilde{Z} = 0$ in (a) and (b) occur due to a lack of divergence in equation~\ref{SHEAR} and the neglect of a short-range excluded-volume potential at the surface. (c) Diagram demonstrating roughly where the far-field approximation is valid and where the near-surface effects need to be considered (extended-chain theory).  \label{Fig3} \vspace{-15pt}}
\end{figure}

In Figure~\ref{Fig2}b we plot the result of equation~\ref{SHEAR} for a shear flow for chains of $2N = 20$, $50$, and $80$ with an effective shear rate of $\dot{\gamma} \tau \sin{2 \theta} = \dot{\gamma} \tau_{eff}= 0.1$. We plot two types of simulations; open symbols indicate simulations where fluctuations are turned off, and filled symbols indicate simulations with thermal fluctuations included. It is clear that, especially in the limit of large $N$, the theory and simulation match well. This plot demonstrates a non-monotonous lift force with two characteristic regimes for this loading profile; the near-surface regime demonstrates an increasing $\tilde{F}_L$ with $\tilde{Z}$, and the far-field regime demonstrates a $\tilde{F}_L \sim \tilde{Z}^{-2}$ decay that is expected from the far-field dumbbell model (plotted as dotted lines in Figure~\ref{Fig2}a). We can rescale this graph for all values of $N$ by rescaling $\tilde{Z} \rightarrow \tilde{Z}/N$ and $\tilde{F}_L \rightarrow \tilde{F}_L/N^2$ to collapse all of the curves for a given value of $\dot{\gamma} \tau_{eff}$ onto a single curve. This is shown in the inset of Figure~\ref{Fig2}. There are deviations between theory and simulation at small values of $\tilde{Z}$ due to the discretized representation of the chain in the simulations as opposed to continuous representation of the chain in the theory. A discretized version of our theory can be developed that only requires a few terms at low $\tilde{Z}$ (see SM) which more accurately reflects the low $\tilde{Z}$ behavior in our simulations, and is shown as a dashed line in Figure~\ref{Fig2}b. We expect that a real chain would more closely resemble the continuous case, as the discretization of the chain is arbitrarily done to develop a convenient simulation model. 

The size and shape of the depletion layer near a surface can now be calculated using equation~\ref{SHEAR}. This is done by describing a potential of mean force $\tilde{U}_L = - \int_{\infty}^{\tilde{Z}}{\tilde{F}_L d \tilde{Z}}$. If the only force acting on the polymer in the $\tilde{Z}$-direction is the lift force, we can write the concentration $c(\tilde{Z})$ of a dilute polymer solution at height $\tilde{Z}$ from the surface as $c(\tilde{Z}) = c(\infty)e^{-\tilde{U}_L}$. We plot sample profiles for our model, for the dumbbell model (where we use both the far-field result and the dumbbell result given by Hoda and Kumar) for shear flow in Figure~\ref{Fig3}a.~\protect\cite{Hoda:2007p7378} There is a striking difference between these two profiles, with drastic differences appearing due to the significantly lower lift forces close to the surface using our elongated chain model rather than the dumbbell model. We also predict a finite concentration at the wall, at least until steric or other short-range wall forces become significant, since the effective lift force potential does not diverge at $\tilde{Z} = 0$. These characteristics have been noticed before in experiments, with depletion layer widths that are smaller than the predicted ones by as much as an order of magnitude, and which are often accompanied by finite surface concentrations of polymer.\protect\cite{Fang:2007p7340} Our model clearly provides an explanation of the origin of the difference between experiment and previous theories, which do not adequately account for near-surface hydrodynamics.

We can also consider flows in a slit, where two walls are present. If the slit height $H$ is on the same order of magnitude as the polymer contour length $Na$, we demonstrate extremely large differences between existing theories and our elongated chain model. Since we have introduced an expression for $\tilde{F}_L(\tilde{Z}, N, \tilde{F}_T)$ that is based off a single image chain, we simply need to consider an infinite series of image chains. For a slit of height $H$, this becomes $\tilde{F}_{L,slit} = \sum_{j = 0}^{\infty} \tilde{F}_L(jH/a+\tilde{Z},N, \tilde{F}_T) -  \tilde{F}_L((j+1)H/a-\tilde{Z},N, \tilde{F}_T) $. We plot a sample distribution function in a slit for our chain elongation model and both the far-field and non-far-field dumbbell models in Figure~\ref{Fig3}b, where the far-field behavior is held constant. While the two dumbbell cases are essentially identical in this regime, there is a marked difference in the distribution function that results from considering the elongated chain geometry.

These depletion layer calculations emphasize where our elongated-chain theory provides a meaningful improvement over dumbbell-based models, which is at high-shear and near-surface conditions. Higher shear rates result in a longer effective chain, which increases the height $\tilde{Z}^*$ where there is a crossover from the near-surface to the far-field regimes. This is demonstrated schematically in Figure~\ref{Fig3}c, which indicates the extent of the near-surface regime as a function of flow rate (shown here in terms of the Weissenberg number $Wi = \dot{\gamma} \tau_Z$, for ease of comparison with traditional literature).

In summary, we have developed an analytical expression for the lift force on a polymer near a surface that significantly improves upon existing theories by accounting for extended geometries. Computer simulations incorporating the hydrodynamic forces were performed and confirmed these analytical expressions, which give rise to the appearance of non-monotonic lift force behavior. This non-monotonic lift force behavior appears in a near-surface, highly extended regime, and it is not captured by a dumbbell theory. Furthermore, our results have important implications for slit flows in microfluidics and polymer adsorption, where the distance of the polymer from the surface and its contour length are on the same order of magnitude.\protect\cite{NETZ} Figure~\ref{Fig3}b suggests that this result could be verified experimentally using the visualization of the cross-channel distribution fluorescently-labelled polymers by applying strong shear flows in narrow channels where this effect is particularly large. Cross slit flows may also provide a tool to directly examine the lift-force on individual polymers.

We acknowledge funding from the National Defense Science and Engineering Graduate Fellowship

\section{Supplemental Information}

\subsection{Hydrodynamic Mobility Tensors}
To describe the effect of a force applied to a bead at $\mathbf{r}_j$ on the surrounding fluid flow at $\mathbf{r}_i$, we define a mobility tensor $\mu_{ij} (\mathbf{r}_i, \mathbf{r}_j)$ that satisfies the equation $\mathbf{v}_i (\mathbf{r}_i ) = \mu_{ij} (\mathbf{r}_i, \mathbf{r}_j) \cdot \mathbf{F}_j (\mathbf{r}_j )$. There are multiple forms for this tensor, depending on the assumptions made and the system of interest. The simplest version, called the Oseen tensor, describes the effect of a single point force on the surrounding fluid flow in an infinite medium:\protect\cite{BLAKE:1974p2245}
\begin{equation}
\mu_{ij,O} (\mathbf{r}_i - \mathbf{r}_j = \mathbf{r}) = \frac{1}{8 \pi \eta r} \left[ \mathbf{I} + \mathbf{\hat{r} \hat{r}}\right]
\end{equation}
where we define $r$ as the magnitude of $\mathbf{r}$ and $\mathbf{\hat{r}} = \mathbf{r}/r$. To consider beads of finite size, we can instead use the multipole expansion of the Oseen tensor, known as the Rotne-Prager-Yamakawa tensor:\protect\cite{ROTNE:1969p5862, Karrila}
\begin{equation}
\frac{\mu_{ij,RPY} (\mathbf{r})}{\mu_0 } = 
 \frac{3a}{4r_{ij}} \left[ \left( 1 + \frac{2a^2}{3r_{ij}^2} \right) \mathbf{I} +  \left( 1 - \frac{2a^2}{r_{ij}^2} \right) \mathbf{\hat{r} \hat{r}}  \right]  
\end{equation}
where $\mu_0 = 1/(6 \pi \eta a)$ is the Stoke's mobility of a bead of radius $a$.

Both the Oseen and Rotne-Prager-Yamakawa tensors assume an infinite medium. If there is a surface present, additional terms need to be added to these tensors such that they account for the no-slip boundary condition. The resulting tensor is called the Blake tensor, and is given (for the Oseen-based Blake tensor) by:\protect\cite{BLAKE:1974p2245}
\begin{eqnarray}
\mu_{ij,B} (\mathbf{r}) = \mu_{ij,O}(\mathbf{r}) - \mu_{ij,O} (\mathbf{r} + \mathbf{r}_{j,Z}) + \nonumber \\ + 2 \mathbf{r}_{j,Z}^2 \mu_{ij,PD} (\mathbf{r} + \mathbf{r}_{j,Z}) - 2 \mathbf{r}_{j,Z} \mu_{ij,SD}(\mathbf{r} + \mathbf{r}_{j,Z})
\end{eqnarray}
where $\mathbf{r}_{j,i,Z} = 0$ at the surface and:
\begin{equation}
\mu_{ij,PD} (\mathbf{r}) = (1-2\delta_{x_i,Z}) \left( \frac{\mathbf{I} - 3  \mathbf{\hat{r} \hat{r}}}{r_{ij}^3}  \right)
\end{equation}
and
\begin{equation}
\mu_{ij,SD} (\mathbf{r}) = \mathbf{r}_{Z} \mu_{ij,PD} (\mathbf{r}) +(1-2\delta_{x_i,Z}) \left[ \frac{\delta_{x_j,Z} \mathbf{r}_{i} - \delta_{x_i,Z} \mathbf{r}_j}{ \mathbf{r}_{ij}^3}\right]
\end{equation}

\subsection{Simulation Methods}
To simulate a single chain in the geometry specified in the paper, we use Brownian Dynamics (BD) simulations. Our polymer is represented by a bead-spring model, which is composed of $2N$ beads $i$ at positions $\mathbf{r}_i$ that are held together by a harmonic potential:
\begin{equation}
\tilde{U}_S = \frac{\tilde{\kappa}}{2} \sum_{i=1}^{2N-1}{(\tilde{r}_{i,i+1}-2)^2}
\end{equation}
 The beads interact with other monomers through a Lennard-Jones potential:
\begin{equation}
\tilde{U}_{LJ} = \tilde{\epsilon} \sum_{i,j}^{2N} \left[ \left( \frac{2}{\tilde{r}_{ij}} \right)^{12} - 2 \left( \frac{2}{\tilde{r}_{ij}} \right)^6 \right]
\end{equation}
There is also a harmonic potential that fixes the polymer at a given height $\tilde{Z}$ from the surface:
\begin{equation}
\tilde{U}_Z = \frac{(\sum_j^{2N}{\mathbf{\tilde{r}_{j,z}}}/(2N) - \tilde{Z})^2}{10}
\end{equation}
where all values designated with a tilde are dimensionless, with distances normalized by the bead radius $a$, energies normalized by $kT$, and times normalized by the characteristic diffusion time $\tau = 6 \pi \eta a^3/(kT)$. $\tilde{r}_{ij}$ is the distance between beads $i$ and $j$, $2N$ is the overall number of beads in the chain, $\tilde{\kappa} = 500$ is the bead-bead spring constant, and $\tilde{\epsilon}$ is a bead interaction parameter that controls the strength of bead-bead attraction. For this paper we use the value $\tilde{\epsilon} = 0.41$, which is typical for a $\theta$- polymer. Beads move through this potential via integration of the Langevin equation:
\begin{equation}
\frac{\partial}{\partial \tilde{t}}\mathbf{\tilde{r}_i} =\mathbf{\tilde{v}_\infty} (\mathbf{\tilde{r}_i)} -   \sum_j \left( \tilde{\mu}_{ij,B} \nabla_{\mathbf{r_j}} \tilde{U}_{tot} (\tilde{t}) + \nabla_{\mathbf{r_j}} \cdot \mathbf{\tilde{D}}_{ij} \right) +  \xi_i (\tilde{t})
\end{equation}
where $\mathbf{\tilde{v}_{\infty} (\tilde{r}_i)}$ is the undisturbed solvent flow profile, $\mu_{ij}$ is the Rotne-Prager-Blake mobility matrix, $\tilde{U}_{tot} = \tilde{U}_S + \tilde{U}_{LJ} + \tilde{U}_{Z}$, $\mathbf{D}_{ij} = k_B T \mu_{ij}$ is the diffusion tensor, and $\xi_i$ is a random velocity that satisfies $\langle \xi_i \rangle = 0$ and $\langle \xi_j (t) \xi_i (t') \rangle = 2k_B T \mu_{ij} \delta (t-t')$. Two cases are considered in this letter for $\mathbf{\tilde{v}_{\infty} (\tilde{r}_i)}$. We represent the force loading corresponding to a shear flow with $\mathbf{\tilde{v}_{\infty, x} (\tilde{r}_i)} = \dot{\gamma} \tau (\mathbf{r_{i,x}} - \sum_j^{2N}{\mathbf{r_{j,x}}}/(2N))$, which captures the linear flow increase away from the center of mass of the chain. The pulling case is given by $\mathbf{\tilde{v}_{\infty, x} (\tilde{r}_i)} = \tilde{F}_0 \delta(i-2N) - \tilde{F}_0 \delta(i)$, which essentially only places a force in the $x$-direction on the first and last beads in a chain. The Langevin equation is discretized by a time step of $\Delta \tilde{t} = 10^{-4}$, and for a given set of conditions $10^7$ simulation steps are used. For a given value of $\tilde{Z}$, the average lift force $\tilde{F}_L$ on a polymer is given by:
\begin{equation}
\tilde{F}_L = \frac{\sum_j^{2N}{\mathbf{\tilde{r}_{j,z}}}/(2N) - \tilde{Z}}{5}
\end{equation}

\subsection{Polymer Dimensions in Shear Flow}
The landmark work by DeGennes on chain behavior in strong flow fields allows us to calculate chain dimensions as a function of shear rate.\protect\cite{DeGennes} We desire a relation that yields an approximate "aspect ratio" of a polymer chain in shear flow that serves as the basis for the respective number of beads that would be used in an analogous bead-spring representation of the chain.

We begin by introducing a series of equations that DeGennes derives from the Peterlin Dumbell formalism:\protect\cite{DEGENNES:1974p2152}
\begin{eqnarray}
1 - EC_{xx} + \dot{\gamma} \tau_Z C_{zx} = 0 \label{DG1} \\
-EC_{xz} + \frac{1}{2} \dot{\gamma} \tau_Z C_{zz} = 0 \label{DG2} \\
EC_{zz} = EC_{yy} = 1
\end{eqnarray}
where $E$ is a chain extension parameter that is $1$ at low extensions and diverges at high extensions, $C_{ij} = 3 \langle r_i r_j \rangle / Na^2$ is a symmetric matrix describing the polymer dimensions, and $\dot{\gamma} \tau_Z$ is the dimensionless shear rate (also known as the Weissenberg number) based on the Zimm relaxation time $\tau_Z$. Equations~\ref{DG1} and~\ref{DG2} can be reorganized to give the result:
\begin{equation}
C_{xx} = \frac{1}{E} + \frac{(\dot{\gamma} \tau_Z)^2}{2E^3}
\end{equation}
The aspect ratio of the chain geometry, which will become $2N$, can be then obtained using the relationship:
\begin{equation}
2N = \left( \frac{C_{xx}}{C_{zz}} \right)^{1/2} = \left[ 1+\frac{(\dot{\gamma} \tau_Z)^2}{2E^2} \right]^{1/2}
\end{equation}
This equation yields intuitive behavior - at $\dot{\gamma} \tau_Z < 1$, the polymer is roughly spherical. As $\dot{\gamma} \tau_Z$ becomes greater than $1$, there becomes a regime of linear scaling of the aspect ratio with the shear rate. This regime flattens out when $E>1$ at high extension, and the value of $2N$ asymptotically approaches the contour length $4N$ of the polymer.

\subsection{Derivation of Lift Force for an Elongated Chain (Pulling Scenario)}

In this letter we have derived the behavior using the kernel $(\partial F_T/ \partial n) \approx 6 \pi \eta a^2 \dot{\gamma} \sin{(2 \theta)} n$, however others could be applicable. For example, if there is an abundance of mass at the chain ends, it may be more appropriate to use a pulling profile $(\partial F_T/\partial n) = F_0 \delta(n - N)$. This is essentially the dumbbell result, only we now integrate over the entire chain contour in the same fashion as before:
\begin{eqnarray}
F_L = \sum_{n=1}^N F_{L,d} \approx \int_0^N F_{L,d} dn 
= \frac{3\tilde{Z}^3}{16} \int_0^N{\left( \frac{\partial F_T}{\partial n} \right) \left[ \frac{1}{\left[ (N-n)^2 +  \tilde{Z}^2 \right]^{3/2}} - \frac{1}{\left[ (N+n)^2 +  \tilde{Z}^2 \right]^{3/2}} \right] dn}
\label{Sum2}
\end{eqnarray}
Performing the integration, we get the result:
\begin{equation}
\tilde{F}_{L,pull} = \frac{3 \tilde{F}_0 \tilde{Z}^3}{16} \left[ \tilde{Z}^{-3} - \frac{1}{\left( 4N^2 + Z^2 \right)^{3/2}}\right]
\end{equation}
This simplifies to the result as $\tilde{Z}$ goes to zero:
\begin{eqnarray}
\tilde{F}_{L,pull} \approx \frac{3 \tilde{F}_0}{16}
\end{eqnarray}
and as $Z$ goes to infinity:
\begin{eqnarray}
\tilde{F}_{L,pull} \approx \frac{9 \tilde{F}_0 N^2}{8 Z^2} 
\end{eqnarray}
We can still define a characteristic height $\tilde{Z}^*$ where there is a transition between these two behaviors. Solving for the maximum of the shear behavior, we obtain the result $\tilde{Z}^* \approx N$. Rearrangement of the result for the pulling case can be plotted as $\tilde{F}$ versus $\tilde{Z}/N$. Results for this are shown in Figure 1 (SM), which is analogous to Figure 2 in the letter.

 \begin{figure}
 \includegraphics[width=75mm]{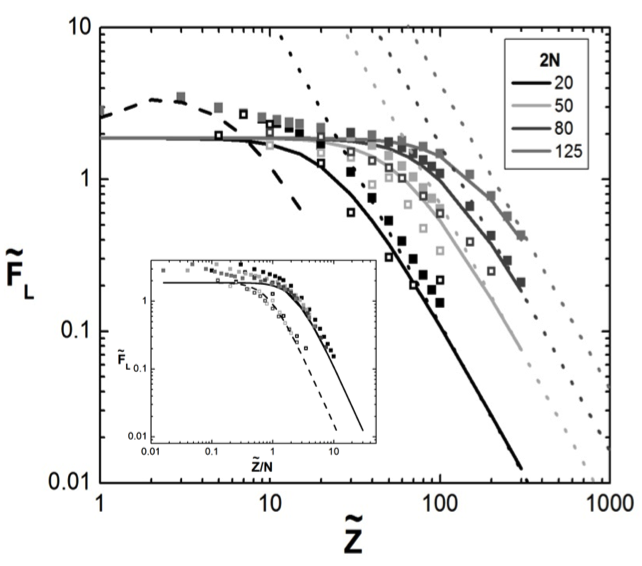}%
 \caption{Graph of $\tilde{F}_L$ versus $\tilde{Z}$ for a shear load profile. Solid lines show theoretical results, dashed lines show the effect of bead-discretization at low values of $\tilde{Z}$, and dotted lines demonstrate the far-field dumbbell results. Simulation data is also shown, with filled symbols representing data without fluctuations and open symbols representing data including fluctuations. These results collapse onto a single curve (inset) using the scaling $\tilde{Z} \rightarrow \tilde{Z}/N$.}
 \end{figure}

\subsection{Low-Z Regime}
The replacement of the summation in equation 2 with an integral is valid if the chain is continuous or if the chain is in the far-field limit. If the chain is discrete and close to the wall, like in our simulations, this approximation does not hold. If we retain the summation in the low-$\tilde{Z}$ limit, we can match our simulation results and demonstrate that this is the reason for the disparity between simulation and theory in Figures 2a and 2b. The discretized version of equation 2 yields the result:
\begin{equation}
F_{L, disc} = \frac{9 \tilde{Z}^3}{16 N} \sum_{n,m = 0}^N{\left[ \frac{(n-m)}{F_{T,n}\left[ (n-m)^2+\tilde{Z}^2\right]^{5/2}} + \frac{(n+m)}{\left[ (n+m)^2+\tilde{Z}^2\right]^{5/2}} \right]} 
\end{equation}
We can simplify this for small $\tilde{Z}$ by considering that the effect  of a given dipole on its surroundings is both antisymmetric and quickly approaches zero. Therefore, beads in the center of the chain will not exert a strong net lift force. Only the beads close to the chain ends contribute non-negligibly to the lift force, which we express as a truncated series:
\begin{equation}
F_{L, disc, shear} \approx \frac{9 \tilde{Z}^3 N \dot{\gamma} \tau}{4}\sum_{i = 0}^l{\left[ \frac{i^2}{\left[ (i)^2+\tilde{Z}^2\right]^{5/2}} \right]}
\end{equation}
for shear and:
\begin{equation}
F_{L, disc, shear} \approx \frac{9 \tilde{Z}^3 F_0}{8}\sum_{i = 0}^l{\left[ \frac{i}{\left[ (i)^2+\tilde{Z}^2\right]^{5/2}} \right]}
\end{equation}
for pulling. $l$ is a small number representing the point at which the summation is truncated (we use $l = 3$). This result agrees with the values obtained for simulation data at small $\tilde{Z}$, as shown by the dotted lines in Figures 2 and 1(SM). This indicates that the main reason for the disparity between theory and simulation is due to the discretization of the chain in the simulations, so therefore we expect that a more realistic chain would more closely agree with the original, continuous theory.

\subsection{Translating From Geometric to Chain Parameters}
The theory presented as such is entirely based on geometric parameters, such as the polymer aspect ratio $N$ and the vertical length scale $a$. In this respect, no particular chain model is assumed and the theory is very general. It is possible to incorporate chain characteristics into such a theory through a change of variables. We take the De Gennes model of a polymer in flow as a simple example, and we use scaling to show how we can obtain the traditional scaling relationship found for the far-field dumbbell model.~\protect\cite{DEGENNES:1974p2152} We use the far-field result for shear flow, $\tilde{F}_{L,shear} \approx \frac{3}{8}  \dot{\gamma} \tau \sin{(2 \theta)} \frac{N^4}{\tilde{Z}^2}$, and make the following scaling replacements:
\begin{eqnarray}
N \sim Wi \\
a \sim \frac{n^{1/2} b}{Wi^{1/2}} \\
\sin{(2 \theta)} \sim Wi^{-1}
\end{eqnarray}
which are appropriate at intermediate values of $Wi$.~\protect\cite{DEGENNES:1974p2152} $n$ is the degree of polymerization, and $b$ is the monomer size. The translation of $\tau$ to $\tau_Z$ is done by noting that $\tau \sim \eta a^3 / (kT)$ and $\tau_Z \sim \eta b^3 n^{3/2} / (kT)$. This leads to the result:
\begin{equation}
\tilde{F}_L \sim \frac{Wi^2 n^{1/2}}{\tilde{Z}^2}
\end{equation}
where all distances are now scaled by $b$. The depletion zone can be calculated by counterbalancing the lift force with the diffusive force pushing the polymer back towards the surface, $F_S$. Since there is a no-flux condition at the surface, we expect the concentration profile to be of the form $c(\tilde{Z}) = c(0) + A \tilde{Z}^2$ where $A$ is an arbitrary constant. The diffusive force is thus given by $\tilde{F}_S \sim \nabla c(\tilde{Z}) \sim -\tilde{Z}$. There is then a critical height $\tilde{Z}^*$ where these forces balance, which is the depletion length:
\begin{equation}
\tilde{Z}^* \sim Wi^{2/3} n^{1/2}
\end{equation}
which is the relationship given in the literature.~\protect\cite{Ma:2005p5986} We add the caveat that these scaling relationships are only appropriate in a finite range of Weissenberg numbers, and is thus not universal. Separate analysis would have to be performed to obtain this sort of relationship under different conditions, however the underlying geometric theory is completely general.

In the letter, order of magnitude calculations are done to relate the geometric parameters to the chain parameters. This is done using the relationship (from the above scaling analysis):
\begin{equation}
\frac{2 Wi^2 n^{1/2} b}{Z} \sim \frac{\dot{\gamma} \tau_{eff} N^4 a}{Z}
\end{equation}
Using the values and relationships $N = 10$, $\dot{\gamma} \tau_{eff} = 0.01$, and $n^{1/2} b/a \sim Wi^{1/2}$, we obtain the result:
\begin{equation}
Wi \sim 5
\end{equation}
This is an order of magnitude result, but it demonstrates that the results presented in this paper are in the relevant shear rate regime to see the effects indicated (lift and depletion).

\bibliographystyle{plainnat}

\begin{thebibliography}{10}

\bibitem{BLAKE:1974p2245}
J~R Blake and A~T Chwang.
\newblock Fundamental singularities of viscous-flow .1. image systems in
  vicinity of a stationary no-slip boundary.
\newblock {\em J Eng Math}, 8(1):23--29, Jan 1974.

\bibitem{CriadoSancho:2000p2666}
M~Criado-Sancho, D~Jou, LF~Del Castillo, and J~Casas-Vazquez.
\newblock Shear induced polymer migration: analysis of the evolution of
  concentration profiles.
\newblock {\em Polymer}, 41(23):8425--8432, Jan 2000.

\bibitem{DEGENNES:1974p2152}
P~G de~Gennes.
\newblock Coil-stretch transition of dilute flexible polymers under ultrahigh
  velocity-gradients.
\newblock {\em J Chem Phys}, 60(12):5030--5042, Jan 1974.

\bibitem{DeGennes}
Pierre~G. de~Gennes.
\newblock {\em Scaling Concepts in Polymer Physics}.
\newblock Cornell Univ. Press.

\bibitem{Fang:2007p7340}
Lin Fang, Chih-Chen Hsieh, and Ronald~G Larson.
\newblock Molecular imaging of shear-induced polymer migration in dilute
  solutions near a surface.
\newblock {\em Macromolecules}, 40(23):8490--8499, Jan 2007.

\bibitem{HernandezOrtiz:2006p7379}
Juan~P Hernandez-Ortiz, Hongbo Ma, Juan~J de~Pablo, and Michael~D Graham.
\newblock Cross-stream-line migration in confined flowing polymer solutions:
  Theory and simulation.
\newblock {\em Phys Fluids}, 18(12):123101, Jan 2006.

\bibitem{Hoda:2007p15}
Nazish Hoda and Satish Kumar.
\newblock Brownian dynamics simulations of polyelectrolyte adsorption in shear
  flow with hydrodynamic interaction.
\newblock {\em J Chem Phys}, 127(23):234902, Jan 2007.

\bibitem{Hoda:2007p16}
Nazish Hoda and Satish Kumar.
\newblock Brownian dynamics simulations of polyelectrolyte adsorption onto
  charged patterned surfaces.
\newblock {\em Langmuir}, 23(4):1741--1751, Jan 2007.

\bibitem{Hoda:2007p7378}
Nazish Hoda and Satish Kumar.
\newblock Kinetic theory of polyelectrolyte adsorption in shear flow.
\newblock {\em J Rheol}, 51(5):799--820, Jan 2007.

\bibitem{Jendrejack:2002p824}
RM~Jendrejack, JJ~de~Pablo, and MD~Graham.
\newblock Stochastic simulations of dna in flow: Dynamics and the effects of
  hydrodynamic interactions.
\newblock {\em J Chem Phys}, 116(17):7752--7759, Jan 2002.

\bibitem{JendrejackdePablo}
RM~Jendrejack, ET~Dimalanta, DC~Schwartz, MD~Graham, and JJ~de~Pablo.
\newblock Dna dynamics in a microchannel.
\newblock {\em Phys. Rev. Lett.}, 91:038102, 2003.

\bibitem{Jendrejack:2004p2674}
RM~Jendrejack, DC~Schwartz, JJ~de~Pablo, and MD~Graham.
\newblock Shear-induced migration in flowing polymer solutions: Simulation of
  long-chain deoxyribose nucleic acid in microchannels.
\newblock {\em J Chem Phys}, 120(5):2513--2529, Jan 2004.

\bibitem{Karrila}
S~Kim and S~J Karrila.
\newblock {\em Hydrodynamics: Principles and Selected Applications}.
\newblock Butterworth-Heinemann.

\bibitem{Ladoux:2000p5497}
B~Ladoux and PS~Doyle.
\newblock Stretching tethered dna chains in shear flow.
\newblock {\em Europhys Lett}, 52(5):511--517, Jan 2000.

\bibitem{Ma:2005p5986}
HB~Ma and MD~Graham.
\newblock Theory of shear-induced migration in dilute polymer solutions near
  solid boundaries.
\newblock {\em Phys Fluids}, 17(8):083103, Jan 2005.

\bibitem{ROTNE:1969p5862}
J~Rotne and S~Prager.
\newblock Variational treatment of hydrodynamic interaction in polymers.
\newblock {\em J Chem Phys}, 50(11):4831--{\&}, Jan 1969.

\bibitem{Sendner:2007p1461}
C~Sendner and R.~R Netz.
\newblock Hydrodynamic lift of a moving nano-rod at a wall.
\newblock {\em Epl-Europhys Lett}, 79(5):58004, Jan 2007.

\bibitem{Sendner:2008p6216}
C~Sendner and R.~R Netz.
\newblock Shear-induced repulsion of a semiflexible polymer from a wall.
\newblock {\em Epl-Europhys Lett}, 81(5):54006, Jan 2008.

\bibitem{NETZ}
A~Serr, C~Sendner, F~M\"{u}ller, TR~Einert, and RR~Netz.
\newblock Single-polymer adsorption in shear: Flattening vs. hydrodynamic lift
  and surface potential corrugation effects.
\newblock {\em Epl-Europhys Lett}, 92:38002, 2010.

\bibitem{Serr:2009p6217}
A~Serr, C~Sendner, and R.R Netz.
\newblock Single polymer adsorption in shear: flattening versus hydrodynamic
  lift and corrugation effects.
\newblock {\em Epl-Europhys Lett}, pages 1--6, Aug 2009.

\bibitem{Watari}
N~Watari, M~Makino, N~Kikuchi, RG~Larson, and M~Doi.
\newblock Simulation of dna motion in a microchannel using stochastic rotation
  dynamics.
\newblock {\em J. Chem. Phys.}, 126:094902, 2007.

\end{thebibliography}

\end{document}